# Aging associated domain evolution in the orthorhombic phase of <001> textured $(K_{0.5}Na_{0.5})Nb_{0.97}Sb_{0.03}O_3$ ceramics

Jianjun Yao*, Jiefang Li, and D. Viehland

Department of Materials Science and Engineering, Virginia Polytechnic Institute and State University, Blacksburg, Virginia 24061, USA

Yunfei Chang and Gary L. Messing

Department of Materials Science and Engineering, Pennsylvania State University, University Park, Pennsylvania 16802, USA

An aging effect due to a domain evolution in $(K_{0.5},Na_{0.5})Nb_{0.97}Sb_{0.03}O_3$ <001> textured ceramics was investigated by piezoresponse force microscopy. We find that such an aging effect is pronounced in the orthorhombic single phase field. A more uniform and finer domain structure was found with aging, which was believed to originate from defect-migration. After poling, large domains with smooth boundaries were found in the aged condition.

*Email: jjyao@vt.edu

In the past decade, environment concerns have inspired research on high-performance lead-free piezoelectrics, which have the potential to replace lead zirconate titanate (or PZT) solid solutions. Since 2004 when Saito et al. proposed the $(K_{0.5}Na_{0.5})NbO_3$ (or KNN) family as such an alternative, [1] much work has been performed to further improve its piezoelectric properties: such as sintering optimization and compositional modifications.[2] A significant increase of the piezoelectric coefficients in <001> textured KNN ceramics compared to randomly oriented ceramics is due to an enhanced ordering of the distribution of grain orientations along <001>. It is known that ferroelectrics with engineered domain configurations exhibit significantly enhanced piezoelectric responses along nonpolar axes.[3,4] The <001> texture guarantees that the application of E is along [001], which is a different crystallographic orientation than that of the {110} polarization vector in the orthorhombic phase of KNN.

KNN undergoes a structural phase transformation sequence on cooling of paraelectric cubic $(C) \xrightarrow{\sim 400°C}$ ferroelectric tetragonal $(T) \xrightarrow{\sim 210°C}$ ferroelectric orthorhombic $(O)$.[5] The $T \rightarrow O$ boundary is known as the polymorphic phase boundary (PPB), designating its difference from the morphotropic phase boundary (MPB) of Pb-based systems. The PPB of KNN solid solutions is nearly independent of x, remaining almost unchanged in temperature for $0 < x < 1$. One of the major differences between PPB and MPB-based ferroelectrics lies in the temperature dependence of the piezoelectric properties. The PPB-based ones generally exhibit large changes in the piezoelectric coefficient near the PPB, however the MPB systems are only weakly temperature dependent and offer a broad temperature range for usage. The observed enhancement of the piezoelectric coefficient in KNN was first attributed to an MPB between orthorhombic and tetragonal phases.[6-8] However, later studies reported a not strong temperature dependence of the electromechanical properties, and the existence of a O→T PPB near room temperature.[9] Recently, X-ray studies of textured KNN ceramics have shown phase coexistence near the PPB over a 30 °C temperature range, where the relative phase volume fractions changed with temperature. Furthermore, increasing E applied along the [001] resulted in a notable increase in the volume fraction of the T phase, effectively shifting the O → T boundary to lower temperatures. An enhancement in the piezoelectric properties was found to accompany this increase in the T volume fraction.[10]

The structure of KNN has been reported to be dependent on stoichiometry. At room temperature, the orthorhombic phase is not the only structure that has been reported. A monoclinic structure has also been found for compositions near the boundary at $x$=0.5.[11] Small substituent concentrations are known to result in the coexistence of orthorhombic and tetragonal phases.[12,13] However, for <001> textured $(K_{0.5}Na_{0.5})Nb_{0.97}Sb_{0.03}O_3$ (KNN-3%Sb) ceramics, x-ray investigations have revealed a single orthorhombic phase field.[14] It has been suggested that nanodomains play a vital role in the high piezoelectric property near the MPB of Pb-based perovskites.[15] Prior studies of domain structures in KNN have revealed that the domain widths are on the order of 20–50 nm. It was found that orthorhombic and tetragonal domain structures coexisted, with domain walls intersecting between nano-domains at 90° which existed adjacent to sub-micron domains having 45° intersections.[13,16] The sub-micron domains were assembled of tetragonal nanodomains having the same (110) twin walls. More recently, it was shown that KNN has various domain morphologies, including featureless domains and striplike domains.[17,18]

The KNN system is known to have a peculiar aging effect, which has been utilized to improve its piezoelectric properties.[12] A defect migration model has been proposed to explain the origin of the said aging.[12,19,20] This model was based on x-ray diffraction data of KNN ceramics, which underwent different thermal treatments. In addition, an aging and re-poling induced enhancement of the piezoelectric properties has been reported for KNN compositions with coexisting orthorhombic (O) and tetragonal (T) phases.[12] However, micro-structural studies concerning domain evolution associated with aging and repoling have not yet been reported. Aging has been proven to be effective in enhancing field induced domain switching in single T phase $BaTiO_3$ crystals,[19,20] via a defect symmetry conforming principle. A symmetry-conforming property could provide a restoring force for reversible domain switching, and consequently a large recoverable electro-strain.[19] .

Here, we present a study of domain evolution driven by an aging of <001> textured $(K_{0.5}Na_{0.5})Nb_{0.97}Sb_{0.03}O_3$ (KNN-3%Sb) ceramics in the orthorhombic single phase field. Aging and repoling increased the piezoelectric properties, as previously reported.[12] We find that such aging results in a more uniform and finer domain structure, which is pronounced for KNN-3%Sb. Large domains with smooth boundaries were observed in this aged condition after poling.

The <001> textured KNN-3%Sb ceramics were prepared by a templated grain growth (TGG) technique developed by Chang et al. [14] Wafers of ceramics were cut into dimensions of $3\times3\times0.3mm^3$, and were subsequently electroded on both surfaces with gold. Aging was achieved as follows: the samples were poled and subsequently heated to 170°C for 30 days. The following properties studies were then performed. Polarization hysteresis (P-E) and strain vs E-field (ε-E) curves were measured at a frequency of 1 Hz using a modified Sawyer-Tower circuit and a linear variable differential transducer (LVDT) driven by a lock-in amplifier (Stanford Research, SR850). Temperature-dependent dielectric constant measurements were done using a multi-frequency LCR meter (HP 4284A) in the temperature range of 30 to 450°C.

Careful investigations of the domain structure were performed by scanning probe microscopy using the piezo-force mode or PFM (DI 3100a, Veeco). For the PFM studies, gold electrodes were deposited on the bottom face of the samples by sputtering and the electroded face was then glued to the sample stage, and the opposite unelectroded surface was scanned by the SPM tip (Veeco, DDESP-10). An ac modulation voltage of 4V (peak to peak) with a frequency of 22 kHz was applied between the conductive tip and the bottom gold electrode.

The P-E hysteresis loops of unaged and aged KNN-3%Sb ceramics are shown in Figure 1(a). A remnant polarization of Pr=20.5 μC/cm$^2$ and a coercive field of Ec=2.1 kV/mm were found in the unaged condition. Aging makes only a slight difference in these values with Pr=21.5μC/cm$^2$ and Ec=1.9 kV/mm. The bipolar ε-E curves for the unaged and aged conditions are shown in Figure 1(b). Aging results in a significant increase of the induced strain：from 0.1% for the unaged condition to 0.16% for the aged one. However, the unipolar ε-E curves revealed a less pronounced increase, as can be seen in Figure 1(c): from 0.079% (unaged) to 0.090% (aged). The reduced strain of the unipolar ε-E response is consistent with changes in the induced polarization, via electrostriction (Q), as

$$\frac{\varepsilon_{unaged}}{\varepsilon_{aged}} = 0.88 \approx \frac{Q}{Q} \frac{P^2_{unaged}}{P^2_{aged}} = 0.91.$$

Thus, the changes in the unipolar strain with aging arise due to those in polarization. However, the much larger changes in the strain of the bipolar ε-E curves with aging must result from domain switching contributions, potentially similar to that of the T domains in the BaTiO$_3$ crystals by the symmetry conforming concept.[19] This would imply significant changes in the domain distribution with aging and/or subsequent poling.

Representative PFM images of the domain morphology for <001> textured KNN-3%Sb with different heat treatments are shown in Figure 2. In the unaged condition (Figs.2a and 2b), the domain morphology did not pose a preferred crystallographic orientation, but rather had irregularly shaped boundaries. The size of these domains varied over a wide range between 0.1 and 5 μm. This abnormal domain distribution may result from residual stresses remaining within the grain structure after sintering. Such domain distributions and morphologies may not be beneficial for enhanced properties, as previous studies near a MPB have shown that the piezoelectric properties significantly increase with decreasing domain size.[13,21-26] The reduced domain size near the boundary is a result of a low anisotropy, allowing for low symmetry structurally bridging phases. A reduced domain size may enable domain redistribution under application of electric field E.

Representative domain morphologies of the aged condition at 170°C after poling are shown in Figures 2c and 2d. Compared with the unaged state, a clear difference can be seen. First, many small domains emerged after aging whose size was on the order of several hundred nanometers. Second, the domain distribution was more uniform, and no notable domain boundaries were found probably due to the small size. These changes are believed to be due to defect migration. [12,19,20] Please note that aging took place at 170 °C, which is in the T phase field and above the O $\rightarrow$ T boundary near 130 °C. Thus, following a symmetry conforming principle, mobile defects would redistribute with time under field developing a tetragonal-like conforming symmetry. Under bipolar drive at temperatures below the PPB, this frozen-in symmetry conforming condition could then enable domain switching between {110} orthorhombic variants, via the structurally bridged tetragonal conforming state.

Next, we studied the effect of poling before and after aging, as shown in Figure 3. Our findings support the above arguments that enhanced bipolar strain results from domain contributions via defect symmetry conformation. In the unaged condition, poling resulted in irregular domain patterns of 2-5μm in size that had rough boundaries (see Figs.3a and 3b). When the aged condition was subsequently poled, the domain morphology had notably different features (Figs. 3c and 3d): macro-domains of 10μm in size became apparent with smooth boundaries. Please note that a single domain state was not observed. It would appear that the more uniform and finer domain structure in the aged condition (Figs. 2c and 2d) is more readily redistributed under E, evolving into larger macro-domain plates. Subsequent application of E along a nonpolar axis at room temperature may then result in broadened domain walls, via the frozen-in defect symmetry conformation along {001}. Such broaden walls could serve as nuclei for domain switching.[27] We note that the domain patterns of <001> textured KNN-3%Sb ceramics observed here did not have any preferred orientations, and that different domain morphologies of KNN ceramics have been reported. [16-18,28] We believe that such differences may result from stoichiometry or sintering conditions.

Finally, we investigated the domain and local structures in the aged condition by transmission electron microscopy (TEM). Typical domain structures by bright field and lattice imaging are shown in Figure 4. The bright field image revealed the presence of fine domains of lengths 100-300nm and widths of 10-30nm. Also, the domain pattern observed by TEM is consistent with that by PFM in the aged condition. Please note that the domain size observed by TEM is often smaller than that by PFM, which is believed to originate from a decreasing domain size with sample thickness. A selected area electron diffraction (SAED) pattern is shown in the inset of Fig.4a. This inset did not reveal diffuse scattering along <001>, as recently reported. [28] To investigate the possibility of A-site cation ordering, high resolution lattice images were obtained for the aged condition (Fig.4b). In this image, we observed that atomic columns were of near uniform intensity, ruling out the possibility of local Na/K positional ordering developing with aging. The inset of Fig.4b shows a power spectrum of the lattice image, whose pattern was consistent with that of the SAED in the inset of Fig.4a

Our findings indicate that the defect symmetry conforming principle can have important consequence on systems containing a PPB boundary between O and T phases. This boundary is driven by temperature rather than composition as for the MPB. Thus, electric fields applied in the high temperature phase develop a defect conforming symmetry consistent with the high temperature phase. On cooling through the PPB, the defect-conforming symmetry of the high temperature phase is then preserved into the lower temperature one. This provides, in a sense, the low temperature phase with a structural link to the high temperature one. Such a link could have significant consequences for domain switching, where fields applied along crystallographic directions different than that of the polarization vector could result in enhanced bipolar strains.

In summary, an aging associated domain evolution in KNN-3%Sb textured ceramics has been investigated. The results revealed pronounced aging effect is in the orthorhombic single phase field. More uniform and finer domain structures were observed, which are believed to originate from a defect-migration. Large sized domains with smooth boundaries were found in the aged condition after poling.

**Acknowledgements –** this work was financially supported by the National Science Foundation (Materials World Network) DMR-0806592, and by the Department of Energy under DE-FG02-07ER46480. J. Yao also would like to thank the financial support from the China Scholarship Council. Authors also give thanks to Dr. W.W. Ge, Dr. Z.K. Xu, Dr. S. Bhattacharyya, and Dr. R. Withers for useful discussions and NCFL in Virginia Tech for the TEM support.

# Reference

1. Y. Saito, H. Takao, T. Tani, T. Nonoyama, K. Takatori, T. Homma, T. Nagaya, and M. Nakamura, Nature (London) 432, 84 (2004).

2. Y. Guo, K. Kakimoto, and H. Ohsato, Appl. Phys. Lett. 85, 4121 (2004).

3. S. E. Park and T. R. Shrout, J. Appl. Phys. 82, 1804 (1997).

4. S. E. Park, S. Wada, L. E. Cross, and T. R. Shrout, J. Appl. Phys. 86, 2746 (1999).

5. D. W. Baker, P. A. Thomas, N. Zhang, and A. M. Glazer, Appl. Phys. Lett. 95, 091903 (2009)

6. E. Hollenstein, M. Davis, D. Damjanovic, N. Setter, Appl. Phys. Lett. 87, 182905 (2005).

7. R. Wang, R. Xie, K. Hanada, K. Matsusaki, H. Bando, M. Itoh, Phys. Status Sol, A 202, R57 (2005).

8. Y. Guo, K. Kakimoto, H. Ohsato, Appl. Phys. Lett. 85, 4121 (2004).

9. S.T. Zhang, R. Xia and T R. Shrout, J. Electroceram. 19,251 (2007).

10. W.W. Ge, J.F. Li, D. Viehland, Y.F. Chang and G. L. Messing, Phys. Rev. B 83, 224110 (2011).

11. J. Tellier, B. Malic, B. Dkhil, D. Jenko, J. Cilensek, and M. Kosec, Solid State Sci. 11, 320 (2009).

12. K. Wang and J.F. Li, Adv. Funct. Mater. 20, 1924 (2010).

13. J. Fu, R. Z. Zuo and Z. K. Xu, Appl. Phys. Lett. 99, 062901 (2011).

14. Y.F. Chang, S.F. Poterala, Z.P. Yang, S. Trolier-Mckinstry and G.L. Messing, Appl. Phys. Lett. 95,232905 (2009).

15. F. Bai, J.F. Li, and D. Viehland, J. Appl. Phys. 97, 054103(2005).

16. R. P. Herber, G. A. Schneider, S. Wagner, and M. J. Hoffmann, Appl. Phys. Lett. 90, 252905 (2007).

17. S.J. Zhang, H. J. Lee, C. Ma, and X.L. Tan, J. Am. Ceram. Soc., 94, 3659 (2011).

18. Y. Zhou, M. Guo, C. Zhang, and M. Zhang, Ceramics International 35, 3253(2009).

19. X. B. Ren, Nat. Mater. 3, 91 (2004).

20. L. X. Zhang, W. Chen, X. B. Ren, Appl. Phys. Lett. 85, 5658 (2004).

21. S. Wada, K. Yako, H, Kakemoto, T. Tsurui, T. Kiguchi, J. Appl. Phys. 98, 014109 (2005).

22. S. J. Zhang, N. P. Sherlock, R. J.Meyer, and T. R. Shrout, Appl. Phys. Lett. 94, 162906 (2009).

23. Y. Xiang, R. Zhang, and W. W. Cao, Appl. Phys. Lett. 96, 092902 (2010).

24. J. J. Yao, L. Yan, W. W. Ge, L. Luo, J. F. Li, D. Viehland, Q. H. Zhang, and H. Luo, Phys. Rev. B 83, 054107 (2011).

25. J.H. Gao, D.Z. Xue,; Y. Wang, D. Wang, L.X. Zhang, H.J. Wu, S.W. Guo, H.X. Bao, C. Zhou, W.F. Liu, S. Sen, G. Xiao, X.B. Ren, Appl. Phys. Lett. 99, 092901 (2011).

26. D.B. Lin, H.J. Lee, S.J. Zhang, F. Li, Z.R. Li, Z. Xu, Script. Mater. 64, 1149 (2011).

27. W.F. Rao, K.W. Xiao, T.L. Chen, Y. Wang, Appl. Phys. Lett. 97, 162901 (2010).

28. Z.G. Yi, Y. Liu, M. A. Carpenter, J. Schiemer, R. L. Withers, Dalton Trans., 40, 5066 (2011).

**List of figures**

Figure 1. Dynamical electromechanical responses for unaged and aged KNN-3%Sb textured ceramics. (a) Polarization hysteresis loops, (b) bipolar strain hysteresis curves, and (c) unipolar strain hysteresis loops.

Figure 2 Domain morphologies of KNN-3%Sb textured ceramics in two different conditions, (a)-(b) unaged and (c)-(d) aged.

Figure 3 Domain morphologies of poled KNN-3%Sb textured ceramics with E=3kV/mm in two different conditions: (a)-(b) unaged, and (c)-(d) aged.

Figure 4 TEM results for KNN-3%Sb textured ceramics: (a) Bright-field image along the [001] zone axis, where the inset shows a SAED; and (b) <001> lattice image, where the inset is the corresponding power spectrum.

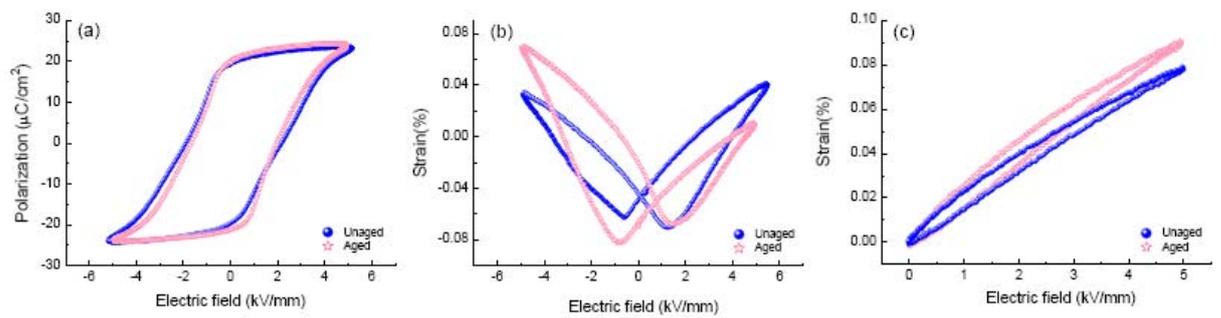

Figure 1

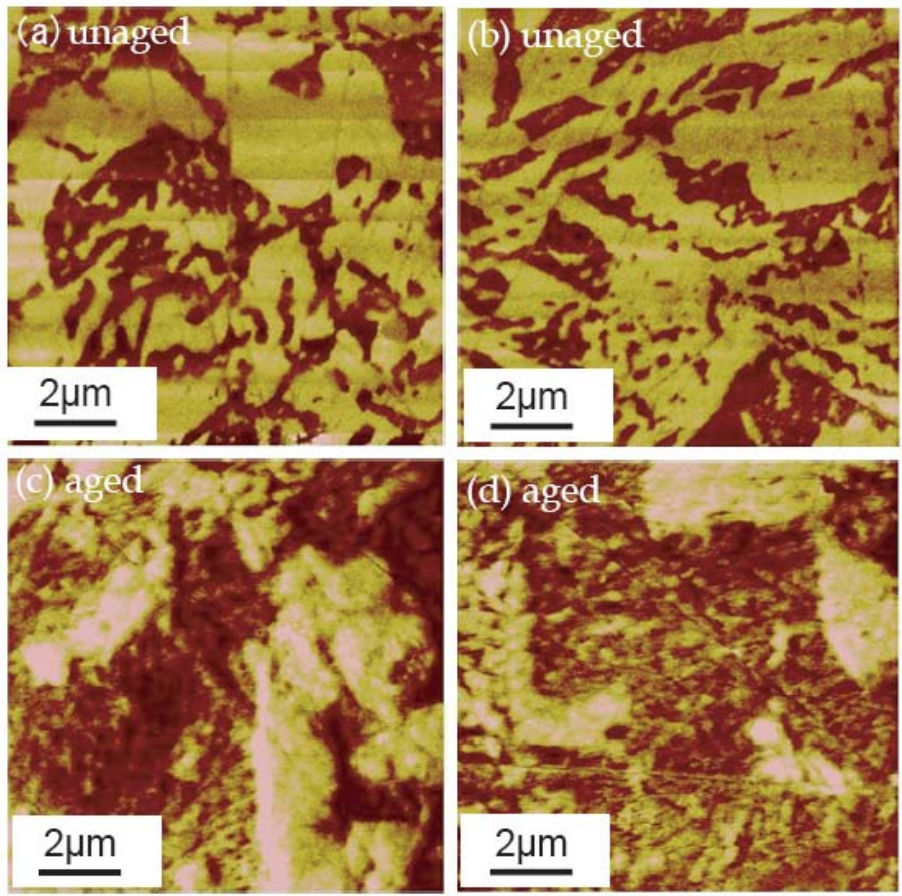

Figure 2

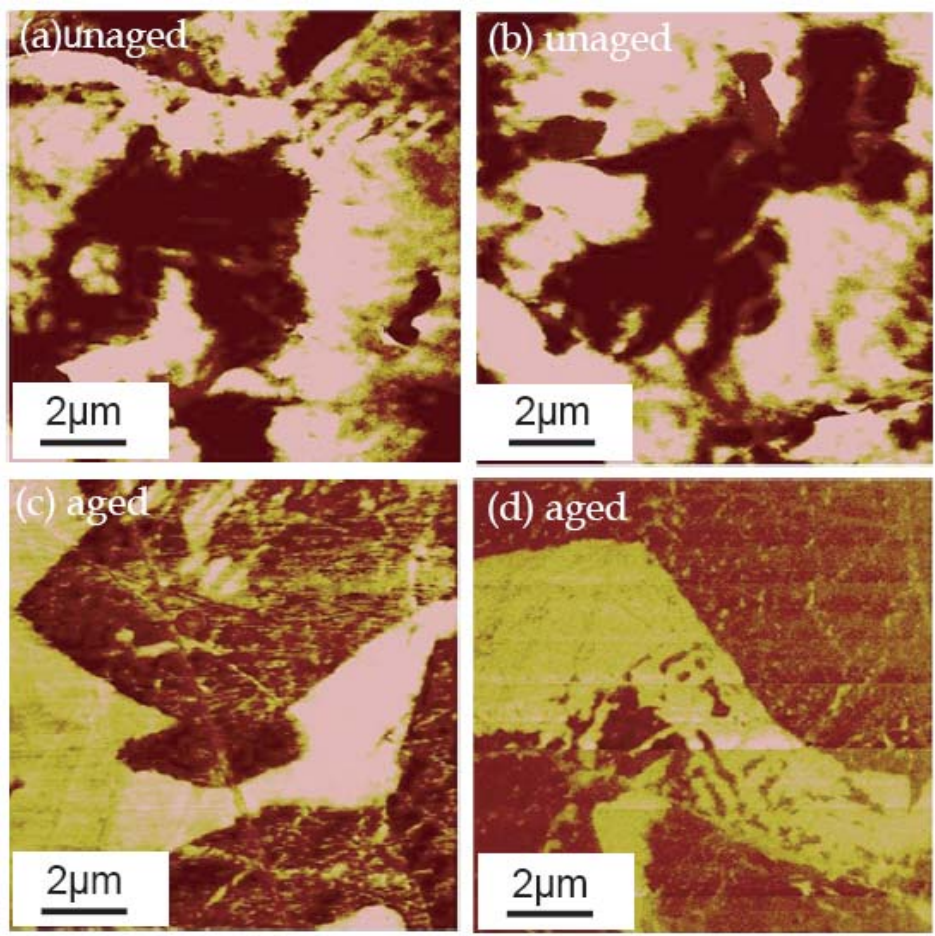

Figure 3

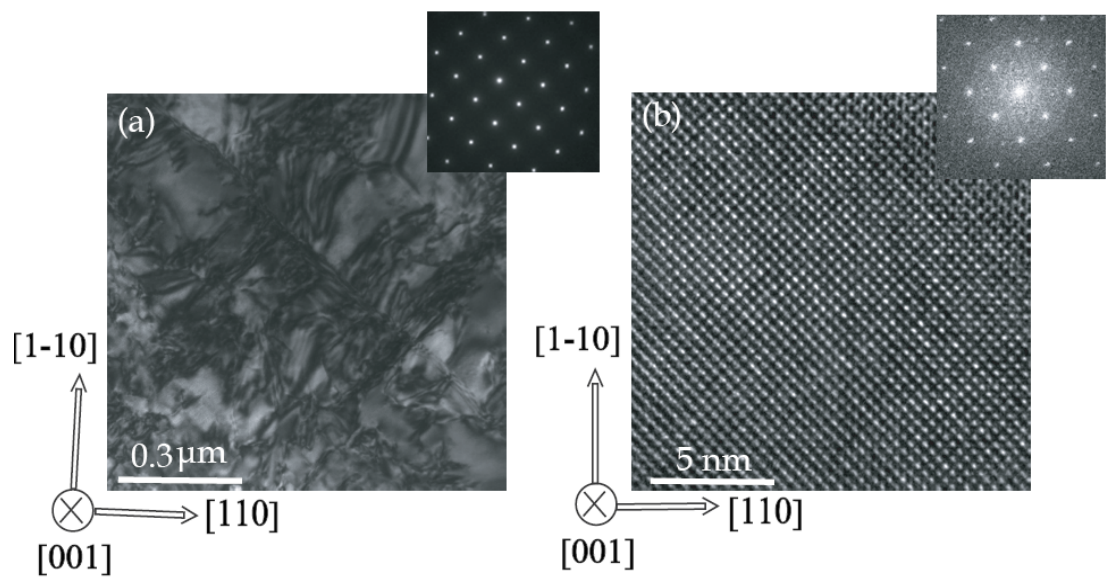

[1-10]

[001] [110]

0.3 μm

[1-10]

[001] [110]

5 nm

Figure 4

*Yao et.al. Aging associated domain evolution in the orthorhombic phase of <001> textured KNN ceramics*